\documentclass[11pt,a4paper]{article}
\pdfoutput=1
\usepackage{amssymb,amsmath,epsf,graphicx}
\usepackage[affil-it]{authblk}

\newcommand{\del}{\ensuremath{\partial}}

\newcommand{\bvphi}{\boldsymbol \varphi}
\newcommand{\bx}{\boldsymbol x}
\newcommand\eq[1]{Eq.~(\ref{#1})}
\newcommand\eqs[2]{Eqs.~(\ref{#1}) and (\ref{#2})}
\newcommand\eqss[3]{Eqs.~(\ref{#1}), (\ref{#2}) and (\ref{#3})}

\newcommand{\ba}{\begin{eqnarray}}
\newcommand{\ea}{\end{eqnarray}}
\newcommand{\beq}{\begin{equation}}
\newcommand{\eeq}{\end{equation}}
\newcommand{\ee}{\end{equation}}
\newcommand{\be}{\begin{equation}}
\newcommand{\es}{\end{split}}
\newcommand{\bs}{\begin{split}}
\newcommand{\eg}{\end{gather}}
\newcommand{\bg}{\begin{gather}}
\newcommand{\eea}{\end{eqnarray}}
\newcommand{\bea}{\begin{eqnarray}}

\date{}
\title{Adiabatic and Isocurvature Perturbation {\em Projections\/} in Multi-Field Inflation}
\author{Chris Gordon
} \affil{Department of Physics and Astronomy, Rutherford Building,
University of Canterbury, Christchurch, New Zealand}
\author{Paul M.\ Saffin}
\affil{School of Physics and Astronomy, University Park, University of Nottingham,
Nottingham NG7 2RD, UK} 

\begin{document}

\maketitle

\begin{abstract}
Current data are in good agreement with the predictions of single field inflation. However, the hemispherical asymmetry, seen in the cosmic microwave background data, may hint at a potential problem. Generalizing to multi-field models may provide one possible explanation. A  useful way of modeling perturbations in multi-field inflation  is to investigate the projection of the  perturbation along and perpendicular to the background fields' trajectory. These correspond to the adiabatic and isocurvature  perturbations. However, it is  important to note that in general there are no corresponding adiabatic and isocurvature fields. The purpose of this article is to highlight the distinction between a field redefinition and a perturbation projection. We  provide a detailed derivation of the evolution of the isocurvature perturbation to show that no assumption of an adiabatic or isocurvature field is needed. We also show how
this evolution equation is consistent with 
the field covariant evolution equations for the isocurvature  perturbation in the flat field space limit.
\end{abstract}

\section{Introduction}
One of the main aims of modern cosmology is to determine if there was a period of inflation in the early Universe. A crucial property of inflation is the number of fields involved. The recent Planck cosmic microwave background (CMB) results, while on the whole being consistent with general single field inflation predictions \cite{Planckinflation,Plancknongauss}, may show the need for multiple fields as there are hints of a possible hemispherical asymmetry in the data \cite{Planckisotropybreaking}.
Similar observations were made in the WMAP data \cite{eri}, but their statistical significance was questioned \cite{lewisbennet}. Confirmation by Planck makes a systematic or foreground explanation less likely. Also, the Planck data shows indications of the modulation extending to smaller scales than was seen in the lower resolution WMAP data. If the primordial fluctuations really are asymmetrical, this isotropy breaking could conceivably result from a large scale mode modulating the smaller scale modes \cite{prunet}. Such a scenario could be realized in a multi-field inflation model \cite{web}. To provide a good fit, the amplitude of the dipole modulation would have to be scale dependent  \cite{mosshoftuft}, especially to accommodate the quasar data \cite{hirata}.

In this article we revisit and clarify a useful method of analyzing inhomogeneity in multi-field inflation models. In Ref.~\cite{gwbm} it was proposed to decompose multi-field perturbations into an adiabatic direction along the  field trajectory ($\delta \sigma$) and an isocurvature direction perpendicular to the field trajectory ($\delta s$). This was found very useful in solving problems related to the preheating of the Universe at the end of inflation and also in understanding how correlated adiabatic and isocurvature perturbations may be generated. Subsequently,  it has also been used extensively in other multi-field inflation investigations. However,  in Ref.~\cite{Paul} it was argued that this procedure leads to an inconsistency. 

Here we show that the apparent inconsistency is due to a conflict of notation within \cite{gwbm}, and that in order to derive the evolution equations of the adiabatic and isocurvature perturbations it is not necessary to assume the existence of an adiabatic and isocurvature field, meaning that the inconsistency does not apply to the results of \cite{gwbm}.

\section{Perturbation Projections vs Field Transformations}
For presentation reasons we will consider the two canonical scalar field (${\bvphi}_1=\phi, {\bvphi}_2=\chi)$ case as was done in \cite{gwbm}.
An extension of the formalism developed in \cite{gwbm} for a larger number of fields and non-canonical couplings can be found in \cite{Nibbelink}.
As in \cite{gwbm}, we consider a background plus a first order perturbation: $\bvphi(t)+\delta\bvphi(\bx,t)$. The perturbation variables satisfy a system of second order linear differential equations. Therefore, new variables defined as linear combinations of the $\phi$ and $\chi$ will also be described by a system of second order linear equations. In \cite{gwbm} it was proposed to look at linear combinations that would project the perturbation parallel ($\delta \sigma$) and perpendicular ($\delta s$) to the background fields' direction of motion in field space, as shown in Figure 1:
\beq
\delta {\sigma }=\sin (\theta ) \delta {\chi }+\cos (\theta ) \delta {\phi },
\label{eq1}
\eeq
\beq
\delta {s }=\cos (\theta ) \delta {\chi }-\sin (\theta ) \delta {\phi },
\label{eq2}
\eeq
where $\theta$ is defined by
\beq
\dot\sigma\equiv \left|\dot{\boldsymbol\bvphi}\right|=\sqrt{\dot\phi^2+\dot\chi^2},
\label{eq3}
\eeq
\beq
\sin({\theta})={\dot{\chi}\over\dot\sigma},\quad \cos(\theta)={\dot{\phi}\over\dot\sigma} \
\label{eq4}
\eeq
and a dot denotes a derivative with respect to time.
\begin{figure}[h]
  \begin{center}
\includegraphics[width=8cm]{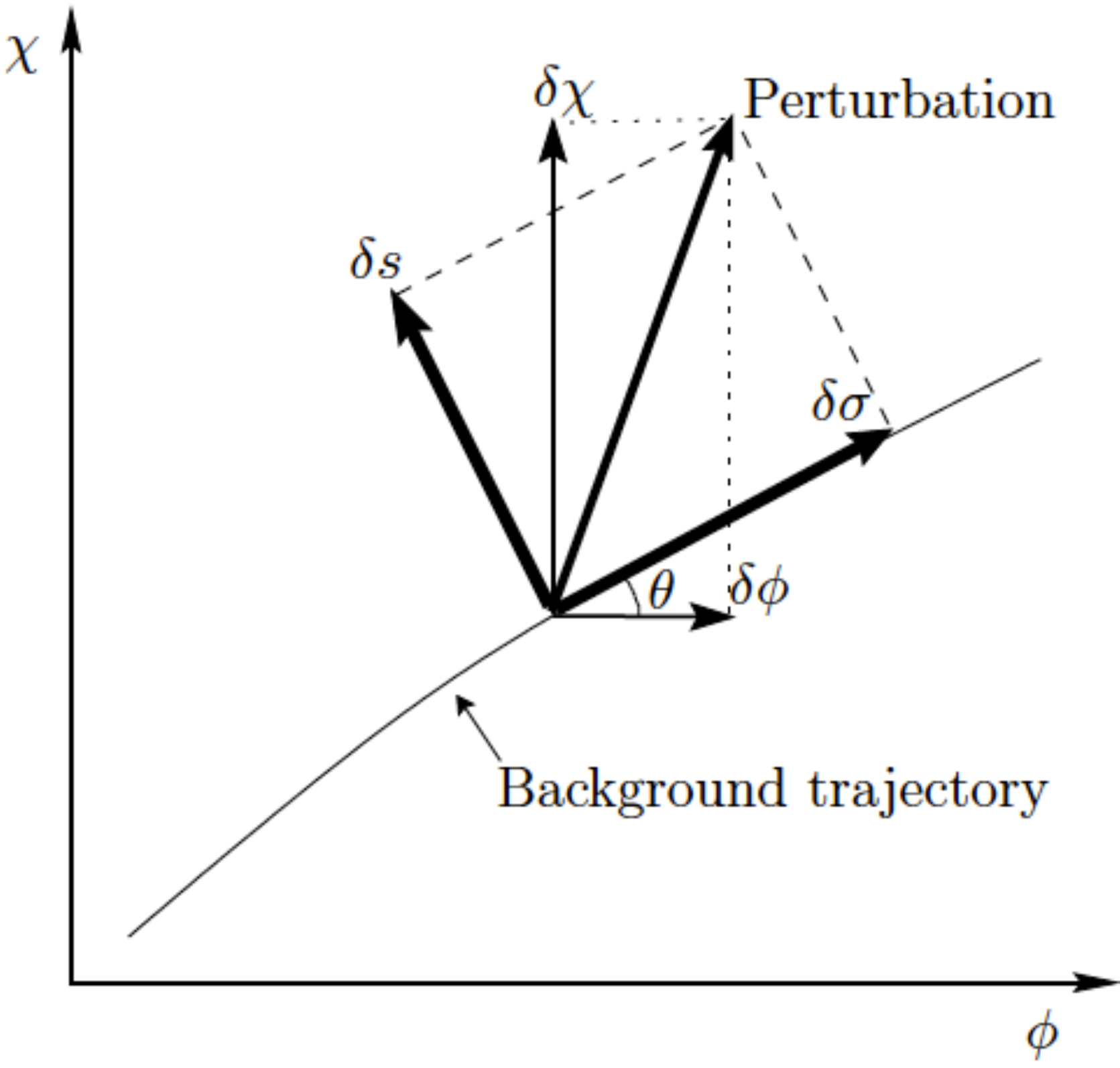}
\caption{
An illustration of the decomposition of an arbitrary perturbation
into an adiabatic ($\delta \sigma$) and isocurvature ($\delta s$)
component. The angle of the tangent to the background trajectory
is denoted by $\theta$. The usual perturbation decomposition,
along the $\phi$ and $\chi$ axes, is also shown.
}
    \label{fig:decomposition}
  \end{center}
\end{figure}

It was shown that $\delta \sigma$ is associated with the adiabatic part of the fields' perturbation and $\delta s$ is the isocurvature (also known as ``entropy'') part of the perturbation. 

It is tempting to think of $\delta \sigma$ and $\delta s$ as actually representing  perturbations in an adiabatic and isocurvature field. However, as recently highlighted in \cite{Paul}, this leads to a contradiction unless $\theta$ is constant. The reason is that if we could write $\sigma=\sigma(\phi,\chi)$ and $s=s(\phi,\chi)$ then working to first order:
\begin{align*}
\delta \sigma&= {\partial \sigma \over \partial \phi}\delta \phi+ {\partial \sigma \over \partial \chi}\delta \chi,\\
\delta s&= {\partial s \over \partial \phi}\delta \phi+ {\partial s \over \partial \chi}\delta \chi.
\end{align*}
Equating these to \eqs{eq1}{eq2} gives
\begin{align}
 {\partial \sigma \over \partial \phi}&=\cos\theta, &{\partial \sigma \over \partial \chi}=\sin\theta,  \label{S1}\\
  {\partial s \over \partial \phi}&=-\sin\theta, &{\partial s \over \partial \chi}=\cos\theta. \label{S2}
\end{align}
Then for $\sigma=\sigma(\phi,\chi)$ and $s=s(\phi,\chi)$ to be a well defined transform we must have ${\partial^2\sigma\over\partial\chi\partial\phi}={\partial^2\sigma\over\partial\phi\partial\chi}$ and ${\partial^2s\over\partial\chi\partial\phi}={\partial^2s\over\partial\phi\partial\chi}$ which from \eqs{S1}{S2} implies that
\[
-\tan\theta{\partial \theta \over \partial \chi}={\partial \theta \over \partial \phi},\quad{\partial \theta \over \partial \chi}=\tan\theta{\partial \theta \over \partial \phi}.
\]
If we combine the above two equations we get the apparently nonsensical result that $\tan^2\theta=-1$. This contradiction can be avoided if we make $\theta$ a constant in \eqs{S1}{S2}.

Although this apparent contradiction may be interesting, it does not have any direct impact on the evolution equations derived for $\delta \sigma$  and $\delta s$ in \cite{gwbm}. The reason being that at no point during that analysis was the assumption made that there was an adiabatic and isocurvature field. All that was done was the projection defining \eqs{eq1}{eq2} were differentiated with respect to time and then the evolution equations of $\delta\bvphi$ were substituted into the result to simplify the equations. 

Some of the terminology and notation in \cite{gwbm} gave the impression that the existence of adiabatic and isocurvature fields was assumed. For example it was stated that subscript notation was used for derivatives, $V_x=\partial V/\partial x$. However, the quantity $V_{ss}\equiv \cos(\theta)^2 V_{\chi\chi}-2 \cos(\theta) V_{\phi\chi} \sin(\theta)+V_{\phi\phi} \sin(\theta)^2$ was also defined, but it was not specified that this was just a definition and not the second partial derivative of the potential with respect to $s$. This form of notation was used as it is suggestive in that if $\theta$ were constant then it would be the correct form for the partial derivative, as then one could actually define an isocurvature field. In fact, any subscripts of the potential which have $\sigma$ or $s$ in them were just meant to be variable definitions, rather than denoting partial derivatives.

\section{Detailed Derivation of Isocurvature Projection Evolution Equation}
In this section we provide a much more detailed derivation of the evolution equation for $\delta s$ that was given in \cite{gwbm}. This is done so as to show explicitly that there is no need to propose the existence of an adiabatic or isocurvature field.

The  Lagrangian density of our two-field system is:
\begin{gather}
{\cal L} = -V({\boldsymbol \bvphi})
 -\frac{1}{2} \sum_{I=1}^{2}  g^{\mu\nu} \partial_\mu \bvphi_I \partial_\nu \bvphi_I
 \,.
\label{lag}
\end{gather}
The field equations, derived from Eq.~(\ref{lag}) for the background
homogeneous fields, are given by the Klein-Gordon equation
\begin{equation}
\label{eq:KG} 
\ddot\bvphi_I + 3H\dot\bvphi_I+ V_{\bvphi_I} = 0\,,
\end{equation}
where $V_{\bvphi_I}\equiv\partial V / \partial \bvphi_I$, a dot denotes the derivative with respect t time, and the Hubble rate, $H$,
in a spatially flat Friedmann-Robertson-Walker (FRW) universe, is
determined by the Friedman equation
\begin{gather}
H^2 = \left(\frac{\dot{a}}{a}\right)^2 =\frac{8\pi G}{3} \left(
V(\bvphi) + \frac{1}{2}  \sum_I (\dot\bvphi_I)^{~2}
 \right)\,, \label{eq:hubble}
\end{gather}
with $a(t)$ the FRW scale factor. This allows us to identify the rate of change of $\theta$, using (\ref{eq3}) and (\ref{eq4}), as
\bea
\dot\theta&=&\frac{1}{\dot\sigma^2}(\dot\chi V_\phi-\dot\phi V_\chi), \nonumber\\
  &=&\frac{\sin\theta}{\dot\sigma}V_\phi-\frac{\cos\theta}{\dot\sigma}V_\chi \nonumber \\
  &=& -{C_{sV}\over \dot\sigma}\, .\label{eq50}
\eea
where $C_{sV} \equiv (\cos\theta) V_\chi - (\sin\theta )V_\phi $.

Scalar field perturbations, with comoving wavenumber $k$, then obey the following  equations in the spatially-flat gauge \cite{taruyanambu}
\be
\ddot{\delta \bvphi_I }+3 H \dot{\delta {\bvphi_I }}+\frac{k^2 }{a^2}\delta {\bvphi_I }+\sum_J B_{IJ }\delta {\bvphi_J }  =0,
\label{perts}
\ee
where
\be
B_{\bvphi_I\bvphi_J}\equiv V_{ \bvphi_I\bvphi_J}-\frac{8  \pi G}{a^3}
   \frac{\text{d}}{\text{dt}}\left(
   \frac{a^3}{H} \dot\bvphi_I \dot{\bvphi}_J
   \right),\label{eq23}
\ee
and $V_{ \bvphi_I\bvphi_J}\equiv \partial^2 V / \partial \bvphi_I\partial \bvphi_J$.
From \eqs{eq1}{eq2}
\bea
\dot{\delta \sigma}&=&\cos(\theta) \dot{\delta {\phi }}+\delta s \dot{\theta }+\dot{\delta \chi } \sin(\theta),\\
\label{eq13}
\dot{\delta s}&=& \cos(\theta) \dot{\delta \chi }-\delta {\sigma } \dot{\theta }-\dot{\delta {\phi }} \sin(\theta), \, 
\label{eq14}
\eea
leading to
\bea
\ddot{\delta \sigma }&=&\cos(\theta) \ddot{\delta \phi }+2 \dot{\delta s} \dot{\theta }+\delta {\sigma } \dot{\theta }^2+\delta s \ddot{\theta }+\ddot{\delta \chi } \sin(\theta),\\
\label{eq15}
\ddot{\delta s}&=& \cos(\theta) \ddot{\delta \chi }-2 \dot{\delta \sigma} \dot{\theta }+\delta s \dot{\theta }^2-\delta {\sigma } \ddot{\theta }-\ddot{\delta \phi } \sin(\theta) \, .
\label{eq16}
\eea
Using $\bvphi_1=\phi$ and $\bvphi_2=\chi$ in \eq{perts} gives
\be
\ddot{\delta s}+3 H \dot{\delta s}+\frac{k^2 }{a^2}\delta s+\dot{\theta } \left(3 H \delta {\sigma }+2 \dot{\delta \sigma}-\delta s \dot{\theta }\right)+\delta {\sigma } \ddot{\theta }+C_s=0,
\label{eq26}
\ee
where
\bea
C_s&\equiv& \cos(\theta) B_{\chi \chi } \delta {\chi }-\sin(\theta) B_{\phi \phi } \delta {\phi }
         +B_{\phi \chi} \left(\cos(\theta) \delta {\phi }-\sin(\theta) \delta {\chi }\right),\\\nonumber
\label{eq27}
   &=&\delta s \left(\sin(\theta)^2 B_{\phi \phi }-\sin(2 \theta ) B_{\phi \chi}+\cos(\theta)^2 B_{\chi \chi }\right)\\
&~&\quad +\delta {\sigma } \left(\cos(2 \theta ) B_{\phi \chi}+\cos(\theta) \sin(\theta) \left[-B_{\phi \phi }+B_{\chi \chi }\right]\right) \, .
\label{eq28}
\eea
We rewrite \eq{eq23} as
\beq
B_{\bvphi_i \bvphi_j }= \left(V_{\bvphi_i \bvphi_j}-\frac{8 \pi  G}{a^3} A_{\bvphi_i \bvphi_j }\right),
\label{eq29}
\eeq
where
\beq
\label{eq30}
A_{\bvphi_i \bvphi_j }\equiv \frac{\rm d}{{\rm d} t}\left(\frac{a^3}{H}
 \dot{\bvphi}_i \dot{\bvphi}_j \right)  \, .
\eeq
From \eqss{eq28}{eq29}{eq30} we find
\beq
C_s=\delta {\sigma } \left(C_{s\sigma V}-\frac{8 \pi  G}{a^3} C_{s\sigma A}\right)+\delta s \left(C_{ssV} - \frac{8 \pi  G}{a^3} C_{ssA} \right),
\label{eq31}
\eeq
where
\bea
\label{eq32}
C_{{s\sigma V}}&\equiv& \left(-V_{\phi\phi}+V_{\chi\chi}\right)  \cos(\theta) \sin(\theta)+V_{\phi\chi} \left(\cos(\theta)^2-\sin(\theta)^2\right),\\
\label{eq33}
C_{ssV}&\equiv& \cos(\theta)^2 V_{\chi\chi}-2 \cos(\theta) V_{\phi\chi} \sin(\theta)+V_{\phi\phi} \sin(\theta)^2,
\eea
and the expressions for $C_{s\sigma A}$ and $C_{ssA}$ are obtained by replacing $V_{\bvphi_1\bvphi_2}$ with $A_{\bvphi_1\bvphi_2}$ in the corresponding expressions for $C_{s\sigma V}$ and $C_{ssV}$. In \cite{gwbm}, instead of notation such as $C_{ssV}$,  $V_{ss}$  was used. We avoid this here so as not to lead to the incorrect impression that we are taking the derivative of the potential with respect to an adiabatic or isocurvature field. This interpretation would only be correct in the constant $\theta$ case. 
Using the background Klein-Gordon equations in \eq{eq30} gives
\be
\label{eq34}
\begin{split}
A_{\bvphi_i \bvphi_j }=& \frac{3 a^2 \dot{a} \dot{\bvphi_i } \dot{\bvphi_j }}{H}-\frac{a^3 \dot{H} \dot{\bvphi_i } \dot{\bvphi_j }}{H^2}+\frac{a^3 \left(-V_{ \bvphi_i }-3 H \dot{\bvphi_i }\right) \dot{\bvphi_j }}{H}\\
&+\frac{a^3 \dot{\bvphi_i } \left(-V_{ \bvphi_j }-3 H \dot{\bvphi_j }\right)}{H} \, .
\end{split}
\eeq
Using the ``$A$" version of \eq{eq32} and \eq{eq33} leads to
\bea
C_{s\sigma A}&=& \frac{a^4 \dot\sigma \left(-\cos(\theta) V_{ \chi }+V_{ \phi } \sin(\theta)\right)}{\dot{a}},\\
\label{eq35}
C_{{ssA}}&=&0,
\label{eq36}
\eea
which may be used in \eq{eq31} so that the evolution equation \eq{eq26} for $\delta s$ becomes
\beq
\begin{split}
\label{eq38}
&\ddot{\delta s}+3 H \dot{\delta s}+ \frac{k^2 \delta s}{a^2}+
\dot{\theta } \left(3 H \delta {\sigma }+2 \dot{\delta \sigma}-\delta s \dot{\theta }\right)+\delta {\sigma } \ddot{\theta }\\
&+\frac{1}{H} \delta {\sigma } \left\{8  \pi G  \cos(\theta ) V_{ \chi } \dot\sigma
+H \cos(2\theta ) V_{\phi\chi} \right. \\
&\left. +\left[H \cos(\theta ) \left(V_{\chi\chi}-V_{\phi\phi}\right)-8 \pi G  V_{ \phi } \dot\sigma\right] \sin(\theta )\right\}  \\
&+\delta s \left(\cos(\theta )^2 V_{\chi\chi}+\sin(\theta ) \left[V_{\phi\phi} \sin(\theta )-2 \cos(\theta ) V_{\phi\chi}\right]\right)=0\, .
\end{split}
\eeq
Upon using \eq {eq4} this may be expressed as
\beq
\label{eq39}
\begin{split}
&\ddot{\delta s}+3 H \dot{\delta s}  \\
&+\delta s \left(\frac{k^2}{a^2}+\cos(\theta )^2 V_{\chi\chi}-\dot{\theta }^2-2 \cos(\theta ) V_{\phi\chi} \sin(\theta )+V_{\phi\phi} \sin(\theta )^2\right) \\
&= -\left(3 H \delta {\sigma }+2 \dot{\delta \sigma}\right) \dot{\theta }+\frac{1}{2} \delta {\sigma }\times  \\
&\Biggl\{ \frac{2 \left[\dot{\chi }^2 \left(\dddot{\phi}\dot{\chi }-2 \ddot{\phi } \ddot{\chi }\right)+\dot{\phi }^2 \left(\dddot{\phi}\dot{\chi }+2 \ddot{\phi } \ddot{\chi }\right)-\dot{\phi }^3 \dddot{\chi}-\dot{\phi } \dot{\chi } \left(2 \ddot{\phi }^2-2 \ddot{\chi }^2+\dot{\chi } \dddot{\chi}\right)\right]}{\dot\sigma^4} 
\\
&
 -2 \cos(2\theta ) V_{\phi\chi}+\frac{16 \pi G   \dot\sigma \left(V_{ \phi } \sin(\theta )-V_{ \chi }\cos(\theta ) \right)}{H} 
  \\&
+\left(V_{\phi\phi}-V_{\chi\chi}\right) \sin(2\theta )\Biggr\} \, .
\end{split}
\eeq
As shown in  \cite{GBW} the three-curvature perturbation, $\zeta$, is related to the
 curvature perturbation in the longitudinal gauge, $\Psi$, and adiabatic, isocurvature perturbations by
\bea
\dot{\zeta }&=&\frac{H}{\dot{H}} \frac{k^2}{a^2} \Psi +2 \frac{H}{\dot\sigma} \dot{\theta } \delta s,\\
\label{eq40}
\zeta &=& \frac{H}{\dot\sigma} \delta {\sigma } \, ,
\label{eq41}
\eea
giving
\beq
\frac{\delta {\sigma } \dot{H}}{\dot\sigma}+\frac{H \dot{\delta \sigma}}{\dot\sigma}-\frac{H \delta {\sigma } \frac{{\rm d} \dot\sigma}{{\rm d}t}}{\dot\sigma^2}=\frac{H k^2 \Psi }{a^2 \dot{H}}+\frac{2 H \delta s \dot{\theta }}{\dot\sigma},
\label{eq43}
\eeq
which allows us to substitute for $\delta\dot\sigma$ in \eq{eq39}. This, along with
\beq
\dot{H}= -4\pi  G \dot\sigma^2 ,
\label{eq47}
\eeq
and the background Klein-Gordon equation, means we can simplify the right-hand-side of \eq{eq39} to
\beq
RHS_{\delta s}= -\frac{\dot{\theta } \left( 8\dot{\theta }\dot\sigma^2 \delta s-\frac{k^2 \Psi \dot\sigma}{a^2  \pi G }\right)}{2 \dot\sigma^2} \, .
\label{eq48}
\eeq

Finally, we can use \eq{eq33} to simplify \eq{eq39} to
\beq
\ddot{\delta s}+3 H \dot{\delta s}+\delta s \left(\frac{k^2}{a^2}+3 \dot{\theta }^2+C_{ssV} \right)=\frac{k^2 \Psi  \dot{\theta }}{2  \pi G   a^2\dot\sigma}.
\label{eq49}
\eeq

\eq{eq49} matches the evolution equation derived in \cite{gwbm}. Therefore, we have explicitly shown how to derive the evolution equation for the isocurvature perturbation projection without having to define an isocurvature or adiabatic field. A similar calculation can be used to derive the evolution equation for the adiabatic perturbation projection. 
\section{Comparison with the Covariant Approach}
Having made explicit the calculation appearing in \cite{gwbm} we should check that the approach of \cite{Paul} gives consistent answers. In that paper the author wrote down the multi-field formalism in a manner that maintained explicit covariance under field re-definitions, whilst also allowing for the field-space to be curved - see also\cite{Gong:2011uw}. In order to apply that formalism to the current case we shall take the field-space to be flat, and we shall use Cartesian co-ordinates $\phi$ and $\chi$ to represent the $\varphi^\alpha$. The proper-speed in field space is denoted $\dot\sigma$ and is given by
\bea
\dot\sigma^2&=&(\dot\varphi,\dot\varphi),
\eea
where the $(\;,\;)$ denotes the field-space inner-product. The adiabatic and isocurvature perturbations were then defined as
\bea
\delta\sigma&=&\frac{(\dot\varphi,\delta\varphi)}{\dot\sigma},\\
\delta S^{\alpha\beta}&=&2\frac{\dot\varphi^{[\alpha}\delta\varphi^{\beta]}}{\dot\sigma}
\eea
where the antisymmetrization is denoted by $[x,y] \equiv \frac{1}{2}
(xy-yx)$.
We note that the appropriate field-perturbation variable identified in \cite{Paul}, i.e. the one that transforms as a tensor under field re-definitions, reduces to $\delta\varphi^\alpha$ in the Cartesian co-ordinates we are using. For the case of two fields there is a single isocurvature mode, and so we identify
\bea
\delta S^{\phi\chi}&=&\delta s.
\eea
Having done that, we work on the $\{\alpha,\beta\}=\{\phi,\chi\}$ component of (3.16) in \cite{Paul}, using our \eqs{eq33}{eq50} to yield
\bea\nonumber
\ddot{\delta s}+3H\dot{\delta s}+\frac{k^2}{a^2} \delta s+(C_{ssV}-\dot\theta^2)\delta s
  &=&2\frac{C_{sV}}{\dot\sigma^2}(\dot\sigma[\delta\dot\sigma-\dot\sigma A]-\ddot\sigma\delta\sigma).\\\label{eq51}
\eea
Here we have used (3.7) of \cite{Paul} for a spatially-flat background geometry, in spatially-flat gauge, to give
\bea
\frac{4\pi G}{H}\dot\sigma^2\delta\sigma&=&\dot\sigma A,
\eea
in order to replace $\delta\sigma$ with the scalar metric perturbation $A$
\footnote{We have switched notation from \cite{Paul}, here we denote metric perturbation variable $\phi$ of \cite{Paul} by $A$, as $\phi$ is being used as a scalar field.}.
We then relate this to the co-moving density perturbation $\epsilon_m$ \cite{Bardeen:1980kt}\cite{gwbm} 
\bea
\epsilon_m&=&\delta\rho-3H\delta q,\\
  &=&(\dot\varphi,\dot{\delta\varphi}-A\dot\varphi)-(\ddot\varphi,\delta\varphi),\\
  &=&\dot\sigma(\dot{\delta\sigma}-A\dot\sigma)-\ddot\sigma\delta\sigma-\frac{2}{\dot\sigma}\delta S^{\alpha\beta}\del_\alpha V\dot\varphi_\beta\, .
\eea
Using (B.7) and (B.8) of \cite{Paul} we find that the gauge invariant intrinsic curvature perturbation $\Psi$ is given by
\bea
\frac{k^2}{a^2}\Psi&=&-4\pi G\epsilon_m,
\eea
and putting all this together in \eq{eq51} gives precisely (\ref{eq49}).

\section{Conclusions}
When studying multi-field inflation models it is often necessary to make use of both  field transformations and perturbation projections \cite{malikwands,koyamawands,ByrnesWands}. For example Byrnes~and~Wands~(2006)~\cite{ByrnesWands} found that the adiabatic and isocurvature perturbations can be correlated at horizon exit and so it is useful to first 
transform to uncorrelated fields at horizon exit and then use a projection to follow the adiabatic and isocurvature components evolution after horizon exit. 
But, as we have highlighted in this article,  it is not necessary to define a field transformation when one wants to derive the evolution equations of a perturbation projection. 

We have also shown that the field-covariant evolution equations and definitions of the adiabatic and isocurvature perturbations in \cite{Paul} tend to the definitions and equations of motion in \cite{gwbm} in the flat field space limit.
 The main conclusion of this article is that whilst some of the terminology and notation in \cite{gwbm} led to the impression that the projection \eqs{eq1}{eq2} were the result of a transformation to an ``isocurvature'' and ``adiabatic'' field, the actual derivation and use of the evolution equations of $\delta \sigma$ and $\delta s$ never made this incorrect assumption. 

\section*{Acknowledgements}
CG thanks David Wands for helpful discussions.


\end{document}